\documentclass[twocolumn,aps,tightenlines,superscriptaddress]{revtex4}
\usepackage{graphicx}
\usepackage{bm}
\usepackage{epsfig}
\newif\ifpdf
\ifx\pdfoutput\undefined
\pdffalse 
\else
\pdfoutput=1 
\pdftrue
\fi

\newcommand{\bea}{\begin{eqnarray}}
\newcommand{\eea}{\end{eqnarray}}
\newcommand{\beq}{\begin{equation}}
\newcommand{\eeq}{\end{equation}}
\newcommand{\bay}{\begin{array}}
\newcommand{\eay}{\end{array}}

\begin{document}
\ifpdf
\DeclareGraphicsExtensions{.pdf, .jpg}
\else
\DeclareGraphicsExtensions{.eps, .jpg}
\fi
\vspace{1.5cm}
\preprint{ \vbox{
\hbox{ZU-TH 13/02} \hbox{hep-ph/0209230}  }}

\vspace{2.0cm}

\title{
$B^0 \to D^{*-}a^+_1$: Chirality tests and resolving an ambiguity in $2\beta + 
\gamma$~\footnote{~ZU-TH 13/02}
} 

\author{Michael Gronau\footnote{~Permanent address: Department of Physics,
Technion, 32000 Haifa, Israel}}
\affiliation{Institut f\"ur Theoretische Physik, Universit\"at Z\"urich,
8057 Z\"urich, Switzerland}

\author{Dan Pirjol}
\affiliation{Department of Physics and Astronomy, 
The Johns Hopkins University, 
3400 North Charles Street, Baltimore, MD 21218}

\author{Daniel Wyler}
\affiliation{Institut f\"ur Theoretische Physik, Universit\"at Z\"urich,
8057 Z\"urich, Switzerland}

\begin{abstract}

We point out that the decays of $B$ mesons into a vector meson and an 
axial-vector meson can distinguish between left and right-handed 
polarized mesons, in contrast to decays into two vector mesons. 
Measurements in $B^0 \to D^{*-}a^+_1$ are proposed for testing 
factorization and the V$-$A structure of the $b \to c$ current, and for 
resolving a discrete ambiguity in $2\beta + \gamma$.    
\end{abstract}
 
\maketitle

The decays of $B$ mesons into two vector mesons ($B \to V_1V_2$), in which each of the 
two vector mesons decays to two particles whose momenta are measured, can be 
used to study the vector meson polarization \cite{DQSTL}. 
These measurements determine separately rates for linear 
polarization states of the two vector mesons, which are either longitudinal ($0$), or 
transverse to their direction of motion and parallel ($\parallel$) or perpendicular 
($\perp$) to one another. Certain interference terms between corresponding weak hadronic 
amplitudes, ${\rm Im}(H^*_{\parallel} H_{\perp}),~{\rm Re}(H^*_0 
H_{\parallel})$ and ${\rm Im}(H^*_0 H_{\perp})$, can also be measured. These studies 
are useful in obtaining information about weak CKM phases \cite{BB}, but leave 
discrete ambiguities in the extracted values.
Polarization measurements in color-allowed $B \to D^*\rho$ decays \cite{D*rho} are 
in agreement with predictions based on factorization and on heavy quark symmetry 
\cite{factor}. Similar measurements in color-suppressed $B \to J/\psi K^*$ \cite{psiK*}, 
showing deviations from factorization, lack a precise quantitative 
theoretical interpretation \cite{fac-psiK*}.

The above quantities are all invariant under the replacement $(H_0, H_{\parallel},
H_{\perp}) \to (H^*_0, H^*_{\parallel}, -H^*_{\perp})$. Consequently, these experiments 
do not distinguish between right and left-handed 
amplitudes, $H_{\pm} = (H_{\parallel} \pm H_{\perp})/\sqrt{2}$. While they
measure $|H_+|^2 + |H_-|^2 = |H_{\parallel}|^2 + |H_{\perp}|^2$, they do not 
determine the sign of $|H_+|^2 -|H_-|^2 = 2{\rm Re}(H^*_{\parallel} H_{\perp})$.
Therefore, the chiralities of the vector mesons and of the weak currents, 
which couple to them in the factorization approximation, cannot be tested. 
Knowledge of the sign of ${\rm Re}(H^*_{\parallel} H_{\perp})$ in $B^0\to J/\psi K^{*0}$, 
is of particular interest. It would have fixed the sign of 
$\cos 2\beta$ occuring in the time-dependent rate of this process \cite{BB},  
thereby resolving a discrete ambiguity in the weak phase $\beta$ \cite{beta}. Measuring 
additional interference terms between dominant 
and small helicity amplitudes in time-dependent 
$B^0 \to D^{*-}\rho^+$ and $B^0 \to D^{*+}\rho^-$ decays would have removed a sign ambiguity 
in $\sin(2\beta + \gamma)$. 

In the present Letter we propose processes of the type $B \to VA$,
involving vector and axial vector mesons, from which the above missing information can 
be obtained. We will use the example of $B^0 \to D^{*-} a^+_1$ to present three 
major results: (a) a way of measuring separately rates for positive and negative 
helicities, (b) a V$-$A test based on factorization and heavy quark symmetry, and 
(c) a way of resolving the sign ambiguity in $\sin(2\beta + \gamma)$. We will also 
comment on resolving an ambiguity in determining $\beta$ in $B^0 \to J/\psi K^{*0}$ 
by studying the process $B \to J/\psi K_1$, where $K_1$ is an axial vector meson.

Let us explain first the necessary condition for a possible distinction between left and 
right-polarized spin one mesons in hadronic $B$ decays in which only final particle momenta
are measured \cite{GGPR}.
The polarization state of a spin one meson is analyzed through its 
subsequent strong or electromagnetic decay. Since the meson chirality ($\propto 
|H_+|^2 -|H_-|^2$) is odd under parity, terms in the decay distribution which are 
proportional to the chirality must involve parity-odd measurables made-up of final 
particle momenta. Two body decays do not permit such pseudoscalar terms. In three 
body decays, on the other hand, one can form a triple product $(\vec p_1 \times \vec p_2) 
\cdot \vec p_3$, involving for instance the momenta of the three final particles in the 
$B$ rest frame.  
Although a triple product in the decay distribution is also odd under time-reversal, it is
well known that it does not require time-reversal violation when there are final state 
interactions. Thus, a chirality measurement can be performed if the decay amplitude 
involves a nontrivial phase due to final state interactions.   

$B$ decays to a vector meson and an axial vector meson, $B\to VA$, in which the axial vector
meson decays to three pseudoscalars $P_1 P_2 P_3$ via the chain $A \to V'P_1,~V'\to P_2 P _3$, 
demonstrate our point. We note that a similar dependence 
on left versus right polarization does not occur in $B \to V_1 V_2$, where one vector 
meson decays in a chain $V_1 \to V' P_1,~V' \to P_2 P_3$ \cite{GGPR}. 
Two processes of the type $B \to VA$ are $B \to D^* a_1(1230), a_1 \to 3\pi$, measured some
time ago \cite{D*a1}, and $B \to J/\psi K_1(1270), K_1\to K 2\pi$, reported recently
\cite{psiK1}. We will focus our attention mainly on the first decay, and 
will make a few comments on the second process.

A large sample of $18000 \pm 1200$ partially reconstructed $B^0 \to D^{*-}a^+_1$ events, 
combining this mode with its charge-conjugate, was reported very recently \cite{BABARa1}
with a branching ratio 
\beq
{\cal B}(B^0 \to D^{*-} a_1^+) = [1.20 \pm 0.07 ({\rm stat}) \pm 0.14 ({\rm syst})] \%~.
\eeq
The $a_1$ was reconstructed via the decay chain $a^+_1 \to \rho^0\pi^+,~
\rho^0\to \pi^+\pi^-$, while the $D^*$ was identified by a slow pion.
We will show how to measure the sign of $|H_+|^2 - |H_-|^2$ in this process.  
We will also calculate separately the values of $|H_+|^2$ and $|H_-|^2$ in the factorization 
and heavy quark approximation and will explain how they can be measured and 
provide a test for V$-$A versus V$+$A. In order to simplify the discussion, we will first 
neglect a very small amplitude from 
$b \to u \bar c d$ which would introduce 
time-dependent effects via $B^0 - \bar B^0$ mixing. 
Subsequently, when discussing time-dependent measurements, we will include this small 
amplitude, showing how to resolve a discrete ambiguity in a measurement of the weak phase
$2\beta + \gamma$.  

The decay amplitude for $B^0 \to D^{*-} a^+_1$ can be written as a sum over polarizations
of weak decay amplitudes $H_i$ describing $B^0 \to D^{*-}a^+_1$, multiplying corresponding 
strong decay amplitudes $A_i$ for $a_1 \to 3\pi$ ($i=0,+,-$),
\beq\label{A}
A(B^0\to D^{*-} \pi^+(p_1)\pi^+(p_2)\pi^-(p_3)) = \sum_{i=0,+,-} H_i A_i~.
\eeq
We assume that the $a_1$ is reconstructed via $a^+_1 \to \rho^0\pi^+,~\rho^0\to \pi^+
\pi^-$, as in \cite{BABARa1} 
\beq
A_i = A_i(a_1^+ \to \rho^0 \pi^+(p_2)) + A_i(a_1^+ \to \rho^0 \pi^+(p_1))~.
\eeq
The two terms correspond to the two possible ways of forming a $\rho$ meson from 
$\pi^+\pi^-$ pairs.
The $a_1\rho\pi$ coupling can be written in terms of two invariant amplitudes,
\beq\label{a1rhopi}
A(a_1(p,\varepsilon)\to \rho(p',\varepsilon') \pi) =
A(\varepsilon\cdot \varepsilon'^*) + B(\varepsilon\cdot p')
(\varepsilon'^*\cdot p)~,
\eeq
where $(p,\varepsilon)$ and $(p',\varepsilon')$ are the momenta and polarization vectors
of the $a_1$ and $\rho$, respectively.
$A$ and $B$ are related to the $S$ and $D$-wave amplitudes through \cite{SD}
\bea
A &=& A_S + \frac{1}{\sqrt2} A_D\,\\
B &=& \left[ -\left(1-\frac{m_\rho}{E_\rho}\right) A_S -
\left( 1 + 2\frac{m_\rho}{E_\rho}\right) \frac{1}{\sqrt2}A_D\right]
\frac{E_\rho}{m_\rho \vec p_\rho\,^2}~,\nonumber
\eea
where the $\rho$ energy and momentum are given in the $a_1$ rest frame.
Since the ratio of $D$ to $S$ amplitudes is quite small ($-0.107 \pm 0.016$
\cite{PDG}), we will neglect the $D$-wave component and use $A_D=0$ 
in the numerical calculation below.

The strong decay amplitude of $a_1 \to 3\pi$ is obtained by
convoluting the $a_1 \to \rho \pi$ amplitude (\ref{a1rhopi}) with the amplitude for 
$\rho^0(\varepsilon') \to \pi^+(p_i)\pi^-(p_j)$, which is proportional to 
$\varepsilon'\cdot(p_i - p_j)$. One finds  
\bea\label{a1amp}
& &A(a_1^+(p,\varepsilon) \to \pi^+(p_1) \pi^+(p_2) \pi^-(p_3)) \propto \\
& &\qquad
C(s_{13}, s_{23}) (\varepsilon\cdot p_1) +(p_1 \leftrightarrow p_2)~,
\nonumber
\eea
where $s_{ij} = (p_i+p_j)^2$ are Dalitz plot variables, and  
\beq\label{C}
C(s_{13}, s_{23}) = [A + B m_{a_1} (E_3-E_2)] B_\rho (s_{23}) + 2A B_\rho (s_{13})~.
\eeq
Here $B_{\rho}(s_{ij}) = (s_{ij} - m^2_{\rho} - im_{\rho}\Gamma_{\rho})^{-1}$, and 
pion energies are given in the $a_1$ rest frame.

It is convenient to express the $B$ meson decay distribution in the rest frame of the
$a_1$ resonance \cite{GGPR}. We introduce a unit vector normal to the decay plane 
of the three pions, $\hat n \equiv (\vec p_1\times \vec p_2)/|\vec p_1\times \vec 
p_2|$, and denote by $\theta$ the angle between $\hat n$ and the direction $\hat z$ opposite
to the $D^*$ (or $B$) momentum. $p_1$ and $p_2$ are by convention \cite{GGPR}
the momenta of the slow and fast positively charged pions, respectively.
The decay amplitude into final hadronic states depends also on two other (Euler) 
angles $\phi$ and $\psi$. $\phi$ is an angle in the $a_1$ decay plane (I), defining 
the direction of one of the three pions (say $p_3$), while $\psi$ defines the line 
of intersection of the $D^*$ decay plane with a plane (II) perpendicular to $\hat z$. 
Both angles are
measured with respect to the line of intersection of the two planes I and II.
With these notations and convention,
squaring the amplitude (\ref{A}) and integrating over $\phi$ and $\psi$,
one finds for the decay distribution in $\theta$
\bea\label{angdist}
& & \int\int \mbox{d} \phi \mbox{d} \psi |A(B\to D^* 3\pi)|^2  \propto
|H_0|^2 \sin^2\theta |\vec J|^2\\
& & + (|H_+|^2 + |H_-|^2)
\frac12 (1 + \cos^2 \theta) |\vec J|^2\nonumber \\
& & + (|H_+|^2 - |H_-|^2) \cos\theta \mbox{Im}[(\vec J\times \vec J^*)\cdot \hat n]
\nonumber~.
\eea
The vector $\vec J$ is defined in the rest frame of the $a_1$ resonance,
\beq\label{J}
\vec J = C(s_{13}, s_{23}) \vec p_1 + C(s_{23}, s_{13}) \vec p_2~.
\eeq
Note that this expression does not depend on the momenta of the $D^*$ decay
products.

A fit to the angular decay distribution (\ref{angdist}) enables separate measurements 
of the three terms $|H_0|^2,~|H_+|^2 + |H_-|^2$ and $|H_+|^2 - |H_-|^2$. In particular, 
one can measure the overall up-down asymmetry of the $D^*$ (or $B$) momentum direction 
with respect to the $a_1$ decay plane,
\beq\label{As}
{\cal A} = \frac{3}{2}\langle \cos\theta\rangle = \left( \frac{3}{2}R \right )
\frac{|H_+|^2 - |H_-|^2}{|H_0|^2+|H_+|^2 + |H_-|^2}~,
\eeq
where $R$ is defined as
\beq\label{R} 
R = \frac12 \frac{\langle \mbox{Im }[\hat n\cdot (\vec J\times \vec J^*)]
\mbox{sgn }(s_{13}-s_{23})\rangle}{\langle |\vec J|^2\rangle}~,
\eeq
and can be computed using Eqs.~(\ref{C}) and (\ref{J}). Integration over 
the entire Dalitz plot yields a negative value $R =-0.158$. 

The three helicity amplitudes $H_{0,\pm}$ in $B^0 \to D^{*-}a^+_1$, can 
be calculated using factorization \cite{BBNS} and heavy quark symmetry \cite{IW}.
For a V$-$A current $\bar c \gamma_\mu (1 - \gamma_5) b$ we find \cite{factor}
\bea\label{H0}
& &H_0 =
-\frac{G_F}{\sqrt2} V_{cb} V^*_{ud} a_1(D^* a_1)
f_{a_1} \sqrt{m_B m_{D^*}} \\
& &\quad \times\frac{m_B-m_{D^*}}{m_{a_1}} (y+1) \xi(y)~,\nonumber\\
\label{H+-}
& &H_\pm =
\frac{G_F}{\sqrt2} V_{cb} V^*_{ud} a_1(D^* a_1)
f_{a_1} \sqrt{m_B m_{D^*}}\\
& &\quad \times [-(y+1) \pm \sqrt{y^2-1}]  \xi(y)~.\nonumber
\eea
Here $\xi(y)$ is the value of the Isgur-Wise function \cite{IW} at $y = 
(m_B^2+m_{D^*}^2-m_{a_1}^2)/(2m_B m_{D^*}) = 1.43$, $f_{a_1}$ is the $a_1$ decay 
constant and $a_1(D^*a_1)$ is a QCD factor which is close to one. The numerical 
values of these factors do not affect the polarization prediction, for which one 
uses the normalization $|H_0|^2 + |H_+|^2 + |H_-|^2 =1$:
\beq\label{Hi}
|H_0|^2 = 0.75~,~~~~ |H_+|^2 = 0.21~,~~~~|H_-|^2 = 0.04~.
\eeq
These values imply an up-down asymmetry ${\cal A} = - 0.042$, which is measurable with
about 5000 $B^0 \to D^{*-}a^+_1$ and $\bar B^0 \to D^{*+}a^-_1$ events. About three times 
as many events were observed in \cite{BABARa1} in partially reconstructed decays, 
which introduces a dilution factor due to uncertainties in the $D^*$ direction. 
This dilution is avoided in fully reconstructed events, for which present statistics seem 
to be sufficient \cite{sharma}.

The predictions (\ref{Hi}) of the Standard Model apply to $B^0$ decays. 
In $\bar B^0$ decays the values of $|H_+|^2$ and $|H_-|^2$ are interchanged. 
We also note that 
for a V$+$A current the $y + 1$ terms in Eqs~(\ref{H0}) and (\ref{H+-}) 
change sign which corresponds to interchanging $|H_+|^2$ and $|H_-|^2$.
This implies a unique signature for a V$-$A current, following from Eq.~(\ref{As})
and (\ref{Hi}) and from the negative value of $R$. Namely, in the $a^+_1$ rest frame the 
$B^0$ and $D^{*-}$ prefer to move in the hemisphere defined by $\vec p({\rm slow}~\pi^+) 
\times \vec p({\rm fast}~\pi^+)$, while in $\bar B^0$ they prefer to move in the opposite 
hemisphere.

Next let us consider the time-dependent rate of $B^0 \to D^{*-}a^+_1$ following from 
interference of two amplitudes \cite{timedep}, $A$ from $\bar b \to 
\bar c u \bar d$ and $\bar A$ from $b \to u \bar c d$ ($f=D^{*-}(3\pi)^+$),
\bea
A \equiv A(B^0\to f) & = & \sum_{i=0,\parallel,\perp} H_i A_i~,\nonumber\\
\bar A \equiv A(\bar B^0\to f) & = & \sum_{i=0,\parallel,\perp} h_i A_i~,
\eea 
where $A_i$ are calculable {\it complex} functions of $\theta,~\psi$ and $\phi$ as explained 
above. This resembles the situation in $B\to D^*\rho$, in which instead of $A_{0,\parallel}$ 
and $A_{\perp}$ one has $g_{0,\parallel}$ and $ig_{\perp}$, respectively, where $g_i$ are 
{\it real} geometric functions \cite{LSS}.
The three coefficients of the time-dependent rate,
\bea\label{Bt}
\Gamma(B^0(t) \to f) & \propto  & (|A|^2 + |\bar A|^2) + 
(|A|^2 - |\bar A|^2)\cos\Delta mt \nonumber \\
& + & 2 {\rm Im}\left( e^{2i\beta} A \bar A^*\right)\sin \Delta mt~,
\eea 
involve bilinear expressions in $H_i$ and $h_i$ multiplying 
calculable functions of the angle variables, which after integration over $\phi$ are 
given by $R_{ij}\equiv (1/2\pi)\int\mbox{d} \phi {\rm Re}(A_iA^*_j)$ and $I_{ij}
\equiv (1/2\pi)\int\mbox{d} \phi {\rm Im}(A_iA^*_j)$, ($i,j=0,\parallel,\perp$):
\bea
R_{00} & = & \frac12\sin^2\theta |\vec J|^2~,~
R_{\parallel\parallel} = \frac12(1 - \cos^2\psi\sin^2\theta)|\vec J|^2~,\nonumber \\
R_{\perp\perp} & = & \frac12(1 - \sin^2\psi\sin^2\theta)|\vec J|^2~,~
R_{0\parallel} = \sin\psi\sin\theta J^2_n~, \nonumber \\
R_{0\perp} & = & \frac{1}{4}\sin\psi\sin 2\theta |\vec J|^2~,~
R_{\parallel\perp} = \cos\theta J^2_n~, \nonumber \\
I_{0\parallel} & = & -\frac{1}{4}\cos\psi\sin 2\theta |\vec J|^2~,~ 
I_{0\perp} = -\cos\psi\sin\theta J^2_n~, \nonumber \\
I_{\parallel\perp} & = & -\frac{1}{4}\sin 2\psi\sin^2\theta |\vec J|^2~,
\eea
where $J^2_n \equiv (1/2)\mbox{Im}[(\vec J\times \vec J^*)\cdot \hat n]$.

The constant and $\cos\Delta mt$ terms in (\ref{Bt}) determine the real and imaginary 
parts of $H_iH^*_j$ (and $h_ih^*_j$) for {\it all} pairs of transversity amplitudes, 
while the coefficient of $\sin \Delta mt$ contain terms 
\bea\label{sin}
& & {\rm Im}\left [e^{2i\beta}(H_i h^*_j A_i A^*_j + H_j h^*_i A_j A^*_i)\right ] \nonumber \\
& = & {\rm Im}\left [e^{2i\beta}(H_i h^*_j + H_j h^*_i) \right ]{\rm Re}(A_iA^*_j) \nonumber \\
& + & {\rm Re}\left [e^{2i\beta}(H_i h^*_j - H_j h^*_i) \right ]{\rm Im}(A_iA^*_j)~,
\eea
which fix both ${\rm Im}[e^{2i\beta}(H_i h^*_j + H_j h^*_i)]$ and 
${\rm Re}[e^{2i\beta}(H_i h^*_j - H_j h^*_i)]$ for {\it all} pairs $i,j$. This situation 
differs from decays into two vector mesons \cite{LSS}, which do not depend on ${\rm 
Im}(H_0H^*_{\parallel}), {\rm Re}(H_0H^*_{\perp})$ and ${\rm Re}(H_{\parallel}H^*_{\perp})$, 
and which contain only the first term in (\ref{sin}) for $i,j=0,\parallel$ and $i=j=\perp$
and the second term for all other values of $i$ and $j$.  
Writing $H_i = |H_i|\exp(i\Delta_i),~h_i = |h_i|\exp(i\delta_i)\exp(-i\gamma)$
where $\Delta_i$ and $\delta_i$ are strong interaction phases, and using the
above and similar information from $B^0(t) \to \bar f$, enables a determination of $2\beta 
+ \gamma$ without having to measure $|h_i|^2$.
While a similar study of $B \to D^*\rho$ leads to a two-fold ambiguity in the sign of 
$\sin(2\beta+\gamma)$, the new terms in (\ref{sin}) occuring in $B \to D^* a_1$ resolve 
the ambiguity. Detailed algebra will be presented elsewhere \cite{GPWPRD}.

We wish to make a few comments on the process $B^0 \to J/\psi K^*$ from which the 
value of $\sin 2\beta$ can be determined \cite{BB}. We will suggest a possible way by 
learning the sign of $\cos 2\beta$ which would remove a two-fold ambiguity in $\beta$ 
\cite{beta}. As mentioned in the introduction, fixing the sign of $\cos 2\beta$ in 
$B^0 \to J/\psi K^*$ requires knowlege of the sign of $|H_+|^2 - |H_-|^2$ in this process. 
A heuristic argument, using the positive helicity of the $\bar s$ quark in $\bar b \to \bar 
c_L c_L \bar s_L$ seems to suggest that this sign is positive in $B^0$ decays and negative in 
$\bar B^0$ decays. This argument may, however, be affected by final state interactions 
which can flip the $\bar s$ quark helicity within the $K^*$. In fact, a nonzero final state
interaction phase ${\rm Arg}(H^*_0 H_{\parallel})$ was measured in $B^0 \to J/\psi K^*$ 
\cite{psiK*}. 
 
While the sign of $|H_+|^2 - |H_-|^2$ in $B \to J/\psi K^*$ cannot be  measured directly, 
a similar quantity can be measured in $B \to J/\psi K_1$, 
where $K_1(1400)$ is an axial vector meson decaying to $K\pi\pi$. A quark spin
argument suggests here too that the sign is positive in $B$ decays and negative in 
$\bar B$ decays. An experimental confirmation that the sign is unaffected by final 
state interactions would be a useful indication, although not an unambiguous proof,  
that this is true also in $B \to J/\psi K^*$. The sign of $|H_+|^2 - |H_-|^2$ in $B 
\to J/\psi K_1$ can be determined by measuring in the $K_1$ rest frame 
an up-down asymmetry of the $J/\psi$ (or $B$) relative to the $K\pi\pi$ decay plane. 
A quantity $R$, defined as in Eq.~(\ref{R}), was calculated recently \cite{GGPR} for 
$K^+_1(1400) \to K^0 \pi^+\pi^0$ and $K^0_1(1400) \to K^+\pi^-\pi^0$ and was found to be 
positive in both processes, $R=+0.22 \pm 0.03$. 
Assuming that in $B \to J/\psi K_1$ one has $|H_+|^2 - |H_-|^2 > 0$, the up-down 
asymmetry of the $J/\psi$ relative to the $K_1$ decay plane is then expected be opposite 
in sign relative to the asymmetry in $B^0 \to D^{*-} a^+_1$ where R was found to be negative. 
This would serve as an indirect measure of the sign of
$|H_+|^2 - |H_-|^2$ in $B \to J/\psi K^*$.

Let us conclude with a few remarks concerning the chiral structure of the $b$ to $c$ 
weak current, for which a test was proposed here in $B^0 \to D^{*-} a^+_1$. In certain 
extensions of the Standard Model, such as a left-right symmetric model, one expects 
deviations from a pure V$-$A structure  \cite{GrWa}. It was suggested \cite{Volo} that 
such a modification could explain a small discrepancy between exclusive and inclusive 
determinations of $|V_{cb}|$. Several earlier tests of the chiral structure of the $b$ 
to $c$ weak current were 
reviewed in \cite{MG}. It was noted \cite{GW2} that measurements of angular correlations 
in semileptonic decays $B\to (D^*\to D\pi) e\bar\nu$ \cite{CLEOV-A} cannot distinguish
between V$-$A and V$+$A couplings, corresponding to $W_L$ and $W_R$ exchange, 
respectively. This ambiguity may, in principle, be removed by studying decays of polarized 
$\Lambda_b$ baryons \cite{Lambda}, such as produced in $e^+ e^-$ annihilation at the 
$Z$ peak. However, the small $\Lambda_b$ polarization measured at LEP \cite{LEP} seems 
to prohibit such a test. The large number of $B^0 \to D^{*-} a^+_1$ events observed
at $e^+e^-$ $B$ factories \cite{BABARa1} provide an opportunity for answering this question.

We thank J. L. Rosner and V. Sharma for useful comments and A. Hardmeier for checking 
a numerical calculation.
This work was supported in part by the Israel Science Foundation founded by the Israel 
Academy of Sciences and Humanities, by the U. S. -- Israel Binational Science Foundation 
through Grant No.\ 98-00237, by the U.S. NSF Grant PHY-9970781 and by the Schweizerischer 
Nationalfonds.

\def \ajp#1#2#3{Am.\ J. Phys.\ {\bf#1}, #2 (#3)}
\def \apny#1#2#3{Ann.\ Phys.\ (N.Y.) {\bf#1}, #2 (#3)}
\def \app#1#2#3{Acta Phys.\ Polonica {\bf#1}, #2 (#3)}
\def \arnps#1#2#3{Ann.\ Rev.\ Nucl.\ Part.\ Sci.\ {\bf#1}, #2 (#3)}
\def \art{and references therein}
\def \cmts#1#2#3{Comments on Nucl.\ Part.\ Phys.\ {\bf#1}, #2 (#3)}
\def \cn{Collaboration}
\def \cp89{{\it CP Violation,} edited by C. Jarlskog (World Scientific,
Singapore, 1989)}
\def \efi{Enrico Fermi Institute Report No.\ }
\def \epjc#1#2#3{Eur.\ Phys.\ J. C {\bf#1}, #2 (#3)}
\def \f79{{\it Proceedings of the 1979 International Symposium on Lepton and
Photon Interactions at High Energies,} Fermilab, August 23-29, 1979, ed. by
T. B. W. Kirk and H. D. I. Abarbanel (Fermi National Accelerator Laboratory,
Batavia, IL, 1979}
\def \hb87{{\it Proceeding of the 1987 International Symposium on Lepton and
Photon Interactions at High Energies,} Hamburg, 1987, ed. by W. Bartel
and R. R\"uckl (Nucl.\ Phys.\ B, Proc.\ Suppl., vol.\ 3) (North-Holland,
Amsterdam, 1988)}
\def \ib{{\it ibid.}~}
\def \ibj#1#2#3{~{\bf#1}, #2 (#3)}
\def \ichep72{{\it Proceedings of the XVI International Conference on High
Energy Physics}, Chicago and Batavia, Illinois, Sept. 6 -- 13, 1972,
edited by J. D. Jackson, A. Roberts, and R. Donaldson (Fermilab, Batavia,
IL, 1972)}
\def \ijmpa#1#2#3{Int.\ J.\ Mod.\ Phys.\ A {\bf#1}, #2 (#3)}
\def \ite{{\it et al.}}
\def \jhep#1#2#3{JHEP {\bf#1}, #2 (#3)}
\def \jpb#1#2#3{J.\ Phys.\ B {\bf#1}, #2 (#3)}
\def \jpg#1#2#3{J.\ Phys.\ G {\bf#1}, #2 (#3)}
\def \lg{{\it Proceedings of the XIXth International Symposium on
Lepton and Photon Interactions,} Stanford, California, August 9--14 1999,
edited by J. Jaros and M. Peskin (World Scientific, Singapore, 2000)}
\def \lkl87{{\it Selected Topics in Electroweak Interactions} (Proceedings of
the Second Lake Louise Institute on New Frontiers in Particle Physics, 15 --
21 February, 1987), edited by J. M. Cameron \ite~(World Scientific, Singapore,
1987)}
\def \kdvs#1#2#3{{Kong.\ Danske Vid.\ Selsk., Matt-fys.\ Medd.} {\bf #1},
No.\ #2 (#3)}
\def \ky85{{\it Proceedings of the International Symposium on Lepton and
Photon Interactions at High Energy,} Kyoto, Aug.~19-24, 1985, edited by M.
Konuma and K. Takahashi (Kyoto Univ., Kyoto, 1985)}
\def \mpla#1#2#3{Mod.\ Phys.\ Lett.\ A {\bf#1}, #2 (#3)}
\def \nat#1#2#3{Nature {\bf#1}, #2 (#3)}
\def \nc#1#2#3{Nuovo Cim.\ {\bf#1}, #2 (#3)}
\def \nima#1#2#3{Nucl.\ Instr.\ Meth. A {\bf#1}, #2 (#3)}
\def \npb#1#2#3{Nucl.\ Phys.\ B~{\bf#1}, #2 (#3)}
\def \os{XXX International Conference on High Energy Physics, 27 July
-- 2 August 2000, Osaka, Japan}
\def \PDG{Particle Data Group, D. E. Groom \ite, \epjc{15}{1}{2000}}
\def \pisma#1#2#3#4{Pis'ma Zh.\ Eksp.\ Teor.\ Fiz.\ {\bf#1} (#3) #2 [JETP
Lett.\ {\bf#1} (#3) #4]}
\def \pl#1#2#3{Phys.\ Lett.\ {\bf#1}, #2  (#3)}
\def \pla#1#2#3{Phys.\ Lett.\ A {\bf#1}, #2 (#3)}
\def \plb#1#2#3{Phys.\ Lett.\ B {\bf#1}, #2 (#3)}
\def \pr#1#2#3{Phys.\ Rev.\ {\bf#1}, #2 (#3)}
\def \prc#1#2#3{Phys.\ Rev.\ C {\bf#1}, #2 (#3)}
\def \prd#1#2#3{Phys.\ Rev.\ D {\bf#1}, #2 (#3)}
\def \prl#1#2#3{Phys.\ Rev.\ Lett.\ {\bf#1}, #2 (#3)}
\def \prp#1#2#3{Phys.\ Rep.\ {\bf#1}, #2 (#3)}
\def \ptp#1#2#3{Prog.\ Theor.\ Phys.\ {\bf#1}, #2 (#3)}
\def \rmp#1#2#3{Rev.\ Mod.\ Phys.\ {\bf#1}, #2 (#3)}
\def \rp#1{~~~~~\ldots\ldots{\rm rp~}{#1}~~~~~}
\def \si90{25th International Conference on High Energy Physics, Singapore,
Aug. 2-8, 1990}
\def \slc87{{\it Proceedings of the Salt Lake City Meeting} (Division of
Particles and Fields, American Physical Society, Salt Lake City, Utah, 1987),
ed. by C. DeTar and J. S. Ball (World Scientific, Singapore, 1987)}
\def \slac89{{\it Proceedings of the XIVth International Symposium on
Lepton and Photon Interactions,} Stanford, California, 1989, edited by M.
Riordan (World Scientific, Singapore, 1990)}
\def \smass82{{\it Proceedings of the 1982 DPF Summer Study on Elementary
Particle Physics and Future Facilities}, Snowmass, Colorado, edited by R.
Donaldson, R. Gustafson, and F. Paige (World Scientific, Singapore, 1982)}
\def \smass90{{\it Research Directions for the Decade} (Proceedings of the
1990 Summer Study on High Energy Physics, June 25--July 13, Snowmass,
Colorado),
edited by E. L. Berger (World Scientific, Singapore, 1992)}
\def \tasi{{\it Testing the Standard Model} (Proceedings of the 1990
Theoretical Advanced Study Institute in Elementary Particle Physics, Boulder,
Colorado, 3--27 June, 1990), edited by M. Cveti\v{c} and P. Langacker
(World Scientific, Singapore, 1991)}
\def \yaf#1#2#3#4{Yad.\ Fiz.\ {\bf#1} (#3) #2 [Sov.\ J.\ Nucl.\ Phys.\
{\bf #1} (#3) #4]}
\def \zhetf#1#2#3#4#5#6{Zh.\ Eksp.\ Teor.\ Fiz.\ {\bf #1}, #2 (#3) [Sov.\
Phys.\ - JETP {\bf #4}, #5] (#6)}
\def \zpc#1#2#3{Zeit.\ Phys.\ C {\bf#1}, #2 (#3)}
\def \zpd#1#2#3{Zeit.\ Phys.\ D {\bf#1}, #2 (#3)}

\end{document}